\documentclass[showpacs,aps,graphicx,twocolumn]{revtex4}
\usepackage{graphicx}

\begin{document}

\title{Bell-state Analysis for Logic Qubits Entanglement}

\author{Yu-Bo Sheng,$^{1}$\footnote{shengyb@njupt.edu.cn} and Lan Zhou,$^2$ }
\address{$^1$ Institute of Signal Processing  Transmission, Nanjing
University of Posts and Telecommunications, Nanjing, 210003,  China\\
 $^2$College of Mathematics \& Physics, Nanjing University of Posts and Telecommunications, Nanjing,
210003, China\\}

\date{\today }
\begin{abstract}
Decoherence is one of the main obstacles  in long-distance quantum communication.
Recently,  the theoretical work of Fr\"{o}wis and W. D\"{u}r (Phys. Rev. Lett. \textbf{106}, 110402 (2011)) and the experiment of Lu \emph{et al.} (Nat. Photon.  \textbf{8}, 364  (2014)) both showed that the logic qubits entanglement say the concatenated
Greenberger-Horne-Zeilinger (C-GHZ) state is more robust under decoherence. In this paper,  we describe a protocol for Bell-state analysis for this logic qubits entanglement. This protocol can also be extended to the multipartite C-GHZ state analysis. Also, we
discuss its application in the quantum teleportation of a unknown logic qubit and in the entanglement swapping of logic Bell states. As the logic qubits entanglement is more robust under decoherence,  our protocol shows that it is possible to realize the long-distance quantum communication based on logic qubits entanglement.
\end{abstract}
\pacs{ 03.67.Pp, 03.67.Mn, 03.67.Hk, 42.50.-p} \maketitle

\section{Introduction}

Quantum entanglement provides a valuable resource for many important applications in quantum communication, quantum computation and quantum
metrology. For example, quantum teleportation \cite{teleportation}, quantum key distribution \cite{qkd}, and other quantum communication protocols  all require
the entanglement to set up the quantum channel. In one-way quantum computation, they require to create the cluster state to perform the task \cite{one-way}.
However, entanglement is generally fragile. In a practical noisy environment, the entanglement  will decoherence and lose the quantum features.
In current quantum information processing, one of the main goals is to protect the entanglement against influences from the uncontrollable environment. In long-distance quantum communication,  the main approach is to protect the
entanglement encoded in the physical qubits directly.  For example, the approaches of quantum repeaters \cite{repeater} and  photon noiseless linear amplification \cite{amplification} are used  to extend
the distance of the entanglement and protect the photon from the photon loss, respectively. The approaches of entanglement purification and concentration are used to extract the maximally entangled states from the degraded entangled states \cite{purification,concentration}. In quantum computation, the main approach is to utilize the quantum correction code, by encoding a single physical quantum state to a logic quantum qubit which contains many physical qubits \cite{errorcorrection1,errorcorrection2}. Therefore, by using redundant encodings together with manipulations and measurements in such a way that the quantum features can be protected.

 Interestingly, the approach of encoding many physical qubits in a logic qubit has also been discussed for logic qubits entanglement \cite{cghz,yan,pan}. Recently,
 Fr\"{o}wis and D\"{u}r studied a class of quantum entangled state which shows similar feature as the  Greenberger-Horne-Zeilinger (GHZ) state,
 but it is more robust than the normal GHZ state in a noisy environment \cite{cghz}.
 It is called concatenated GHZ (C-GHZ) state of the form
 \begin{eqnarray}
|\Phi_{1}^{\pm}\rangle_{N,m}=\frac{1}{\sqrt{2}}(|GHZ^{+}_{m}\rangle^{\otimes N} \pm |GHZ^{-}_{m}\rangle^{\otimes N}),\label{logic}
\end{eqnarray}
with $|GHZ^{\pm}_{m}\rangle=\frac{1}{\sqrt{2}}(|0\rangle^{\otimes m}\pm|1\rangle^{\otimes m})$.  The degree of such logic qubits entanglement decreases polynomially with particle number in $N$ and $m$. Ding \emph{et al.} described a way of creating the C-GHZ state with cross-Kerr nonlinearity \cite{yan}. Lu \emph{et al.}
reported the experimental realization of a C-GHZ state in optical system with $m=2$ and $N=3$ \cite{pan}. They also demonstrated that the C-GHZ state is more robust than the conventional GHZ state.

On one hand, up to now,  though  several groups discussed the C-GHZ state in both theory and experiment, such logic entangled state has not been studied in current entanglement based quantum communication.
On the other hand,   as the C-GHZ state is more robust, setting up the entanglement channel with logic qubits entanglement rather than the physical qubit entanglement may be an alternative way to resit the noisy environment.
 In this paper,  we will discuss one of the most important two-qubit measurement, say the Bell-state measurement \cite{bellanalysis}. The Bell-state measurement enables many important applications in quantum information processing,  such as teleportation \cite{teleportation}, quantum key distribution \cite{qkd}, and so on \cite{repeater}. Different from the previous Bell-state analysis, we describe the logic Bell-state analysis (LBSA).  Here, the logic Bell state is the state in Eq. (\ref{logic}) with $N=m=2$. It is shown that such state can be deterministic distinguished with the help of controlled-not (CNOT) gate. Moreover, the approach for LBSA can also be extended to distinguish the C-GHZ state with arbitrary $N$ and $m$.

 This paper is organized as follows: In Sec. II, we will describe the approach of the LBSA. In Sec. III, we  will extend this approach for
 distinguishing arbitrary C-GHZ state. Our protocol reveals that the logic Bell-State and the C-GHZ state analysis  can be simplified to the conventional
Bell-state  and GHZ analysis, respectively. In Sec. IV, we will present a discussion. It is shown that, with the help of LBSA, we can teleportate an unknown logic qubit. We also show that
we can perform the complete logic entanglement swapping. In Sec. V, we will  make a conclusion.

\section{Logic Bell-state analysis}

The four logic Bell states can be described as
\begin{eqnarray}
|\Phi^{\pm}\rangle_{AB}=\frac{1}{\sqrt{2}}(|\phi^{+}\rangle_{A}|\phi^{+}\rangle_{B}\pm|\phi^{-}\rangle_{A}|\phi^{-}\rangle_{B}),\nonumber\\
|\Psi^{\pm}\rangle_{AB}=\frac{1}{\sqrt{2}}(|\phi^{+}\rangle_{A}|\phi^{-}\rangle_{B}\pm|\phi^{-}\rangle_{A}|\phi^{+}\rangle_{B}).
\end{eqnarray}
Here
\begin{eqnarray}
|\phi^{\pm}\rangle=\frac{1}{\sqrt{2}}(|0\rangle|0\rangle\pm|1\rangle|1\rangle),\nonumber\\
|\psi^{\pm}\rangle=\frac{1}{\sqrt{2}}(|0\rangle|1\rangle\pm|1\rangle|0\rangle),
\end{eqnarray}
with $|0\rangle$ and $|1\rangle$ are the physical qubit, respectively.
$|\Phi^{+}\rangle_{AB}$ essentially is the state with $m=N=2$ in Eq. (\ref{logic}). In Fig. 1, four physical qubits  comprise two logic qubits A and B. The four physical qubits are in the spatial mode a$_{1}$, a$_{2}$, b$_{1}$ and b$_{2}$ respectively.\\

\begin{figure}[!h]
\begin{center}
\includegraphics[width=6cm,angle=0]{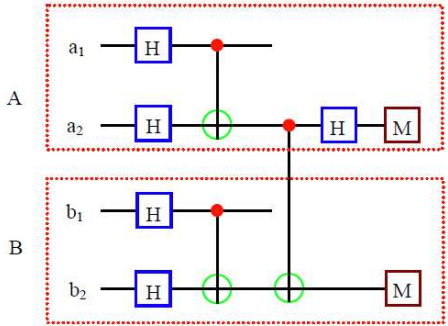}
\caption{Schematic diagram of the logic Bell-state analysis. $H$ represents the Hadamard operation and $M$ represents the measurement in the basis $\{|0\rangle, |1\rangle\}$.}
\end{center}
\end{figure}
As shown in Fig. 1, we first perform a Hadamard operation (H) on each qubit. The Hadamard operation will
make $|0\rangle\rightarrow|+\rangle=\frac{1}{\sqrt{2}}(|0\rangle+|1\rangle)$, and $|1\rangle\rightarrow|-\rangle=\frac{1}{\sqrt{2}}(|0\rangle-|1\rangle)$. They will make
$|\phi^{+}\rangle$ do not change, but $|\phi^{-}\rangle$  become $|\psi^{+}\rangle$.
After performing the four Hadamard operations, the four logic Bell states can be rewritten as
\begin{eqnarray}
|\Phi^{\pm}\rangle_{AB}=\frac{1}{\sqrt{2}}(|\phi^{+}\rangle_{A}|\phi^{+}\rangle_{B}\pm|\psi^{+}\rangle_{A}|\psi^{+}\rangle_{B}),\nonumber\\
|\Psi^{\pm}\rangle_{AB}=\frac{1}{\sqrt{2}}(|\phi^{+}\rangle_{A}|\psi^{+}\rangle_{B}\pm|\psi^{-}\rangle_{A}|\phi^{+}\rangle_{B}).\label{bell}
\end{eqnarray}
  After both performing the CNOT operations, the states $|\Phi^{\pm}\rangle_{AB}$ will evolve as
\begin{eqnarray}
&&|\Phi^{\pm}\rangle_{AB}=\frac{1}{\sqrt{2}}(|\phi^{+}\rangle_{A}|\phi^{+}\rangle_{B}\pm|\psi^{+}\rangle_{A}|\psi^{+}\rangle_{B}),\nonumber\\
&=&\frac{1}{\sqrt{2}}[\frac{1}{\sqrt{2}}(|0\rangle_{a_{1}}|0\rangle_{a_{2}}+|1\rangle_{a_{1}}|1\rangle_{a_{2}})
\otimes\frac{1}{\sqrt{2}}(|0\rangle_{b_{1}}|0\rangle_{b_{2}}\nonumber\\
&+&|1\rangle_{b_{1}}|1\rangle_{b_{2}})
\pm\frac{1}{\sqrt{2}}(|0\rangle_{a_{1}}|1\rangle_{a_{2}}+|1\rangle_{a_{1}}|0\rangle_{a_{2}})\nonumber\\
&\otimes&\frac{1}{\sqrt{2}}(|0\rangle_{b_{1}}|1\rangle_{b_{2}}
+|1\rangle_{b_{1}}|0\rangle_{b_{2}}]\nonumber\\
&\rightarrow&\frac{1}{\sqrt{2}}[\frac{1}{\sqrt{2}}(|0\rangle_{a_{1}}|0\rangle_{a_{2}}+|1\rangle_{a_{1}}|0\rangle_{a_{2}})
\otimes\frac{1}{\sqrt{2}}(|0\rangle_{b_{1}}|0\rangle_{b_{2}}\nonumber\\
&+&|1\rangle_{b_{1}}|0\rangle_{b_{2}})
\pm\frac{1}{\sqrt{2}}(|0\rangle_{a_{1}}|1\rangle_{a_{2}}+|1\rangle_{a_{1}}|1\rangle_{a_{2}})\nonumber\\
&\otimes&\frac{1}{\sqrt{2}}(|0\rangle_{b_{1}}|1\rangle_{b_{2}}+|1\rangle_{b_{1}}|1\rangle_{b_{2}}]\nonumber\\
&=&|+\rangle_{a_{1}}|+\rangle_{b_{1}}\otimes\frac{1}{\sqrt{2}}(|0\rangle_{a_{2}}|0\rangle_{b_{2}}\pm|1\rangle_{a_{2}}|1\rangle_{b_{2}}).\label{evolve1}
\end{eqnarray}
 Similarly, the states $|\Psi^{\pm}\rangle_{AB}$ will evolve as
 \begin{eqnarray}
&&|\Psi^{\pm}\rangle_{AB}=\frac{1}{\sqrt{2}}(|\phi^{+}\rangle_{A}|\psi^{+}\rangle_{B}\pm|\psi^{-}\rangle_{A}|\phi^{+}\rangle_{B})\nonumber\\
&\rightarrow&|+\rangle_{a_{1}}|+\rangle_{b_{1}}\otimes\frac{1}{\sqrt{2}}(|0\rangle_{a_{2}}|1\rangle_{b_{2}}\pm|1\rangle_{a_{2}}|0\rangle_{b_{2}}).\label{evolve2}
\end{eqnarray}
From Eqs. (\ref{evolve1}) and (\ref{evolve2}), one can find that the qubits a$_{1}$ and b$_{1}$ disentangle with the qubits a$_{2}$ and b$_{2}$.
Interestingly, by removing the qubits a$_{1}$ and b$_{1}$, the remained qubits a$_{2}$ and b$_{2}$ essentially are the standard Bell states shown in Eq. (\ref{bell}).
That is the $|\Phi^{\pm}\rangle_{AB}$ become $|\phi^{\pm}\rangle_{AB}$ and $|\Psi^{\pm}\rangle_{AB}$ become $|\psi^{\pm}\rangle_{AB}$, respectively.
The standard Bell states can be well distinguished with the CNOT gate and the Hadamard gate.
From Fig. 1, the qubits a$_{2}$ and b$_{2}$ pass through the CNOT gate in a second time,
and $|\phi^{\pm}\rangle_{AB}$ and $|\psi^{\pm}\rangle_{AB}$ will become
\begin{eqnarray}
|\phi^{\pm}\rangle_{AB}&=&\frac{1}{\sqrt{2}}(|0\rangle_{a_{2}}|0\rangle_{b_{2}}\pm|1\rangle_{a_{2}}|1\rangle_{b2})\nonumber\\
&\rightarrow&\frac{1}{\sqrt{2}}(|0\rangle_{a_{2}}\pm|1\rangle_{a_{2}})|0\rangle_{b2},\nonumber\\
|\psi^{\pm}\rangle_{AB}&=&\frac{1}{\sqrt{2}}(|0\rangle_{a_{2}}|1\rangle_{b_{2}}\pm|1\rangle_{a_{2}}|0\rangle_{b2})\nonumber\\
&\rightarrow&\frac{1}{\sqrt{2}}(|0\rangle_{a_{2}}\pm|1\rangle_{a_{2}})|1\rangle_{b2}.
\end{eqnarray}
Finally, after the qubit a$_{2}$ passing through the Hadamard gate, both qubits are measured in the basis $\{|0\rangle,|1\rangle\}$.
If the measurement result is $|0\rangle|0\rangle$, the original state is $|\Phi^{+}\rangle_{AB}$. If the measurement result is
 $|1\rangle|0\rangle$, the original state is  $|\Phi^{-}\rangle_{AB}$. On the other hand, if the measurement result is  $|0\rangle|1\rangle$
 or  $|1\rangle|1\rangle$, the original state is $|\Psi^{+}\rangle_{AB}$ or $|\Psi^{-}\rangle_{AB}$, respectively. In this way, the LBSA is
 completed.

\section{Entanglement analysis for arbitrary concatenated entangled state}
In this section, we show that the way of distinguishing the LBSA can be used to
analysis the arbitrary C-GHZ state of the form
\begin{eqnarray}
|\Phi^{\pm}_{1}\rangle_{N,m}&=&\frac{1}{\sqrt{2}}(|GHZ^{+}_{m}\rangle^{\otimes N} \pm |GHZ^{-}_{m}\rangle^{\otimes N}),\nonumber\\
|\Phi^{\pm}_{2}\rangle_{N,m}&=&\frac{1}{\sqrt{2}}(|GHZ^{-}_{m}\rangle|GHZ^{+}_{m}\rangle^{\otimes N-1}\nonumber\\
&\pm& |GHZ^{+}_{m}\rangle|GHZ^{-}_{m}\rangle^{\otimes N-1}),\nonumber\\
|\Phi^{\pm}_{3}\rangle_{N,m}&=&\frac{1}{\sqrt{2}}(|GHZ^{+}_{m}\rangle|GHZ^{-}_{m}\rangle|GHZ^{+}_{m}\rangle^{\otimes N-2}\nonumber\\
&\pm& |GHZ^{-}_{m}\rangle|GHZ^{+}_{m}\rangle|GHZ^{-}_{m}\rangle^{\otimes N-2}),\nonumber\\
&\cdots&\nonumber\\
|\Phi^{\pm}_{2^{N-1}}\rangle_{N,m}&=&\frac{1}{\sqrt{2}}(|GHZ^{+}_{m}\rangle^{\otimes N-1}|GHZ^{-}_{m}\rangle\nonumber\\
&\pm&|GHZ^{-}_{m}\rangle^{\otimes N-1}|GHZ^{+}_{m}\rangle).\label{multi1}
\end{eqnarray}
From Eq. (\ref{multi1}), the logic qubits denote as $|GHZ^{\pm}_{m}\rangle$.
The C-GHZ state analysis for the states in Eq. (\ref{multi1}) can be simplified to distinguish the states
\begin{eqnarray}
&&|\Phi^{\pm}_{1}\rangle_{N,2}=\frac{1}{\sqrt{2}}(|\phi^{+}\rangle^{\otimes N}\pm|\phi^{-}\rangle^{\otimes N}),\nonumber\\
&&|\Phi^{\pm}_{2}\rangle_{N,2}=\frac{1}{\sqrt{2}}(|\phi^{-}\rangle|\phi^{+}\rangle^{\otimes N-1}\pm|\phi^{+}\rangle|\phi^{-}\rangle^{\otimes N-1}),\nonumber\\
&&\cdots\nonumber\\
&&|\Phi^{\pm}_{2^{N-1}}\rangle_{N,2}=\frac{1}{\sqrt{2}}(|\phi^{+}\rangle^{\otimes N-1}|\phi^{-}\rangle\pm|\phi^{-}\rangle^{\otimes N-1}|\phi^{+}\rangle.\nonumber\\\label{multi2}
\end{eqnarray}
In C-GHZ state analysis, we only need to distinguish the states in the logic qubit level, and  do not need to care about
the exact information in each logic qubit.
\begin{figure}[!h]
\begin{center}
\includegraphics[width=6cm,angle=0]{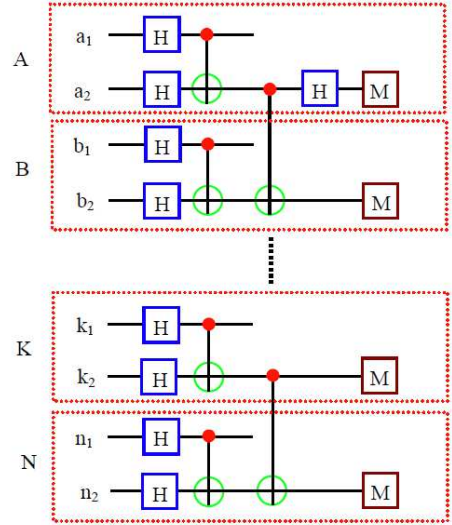}
\caption{Schematic diagram of the arbitrary C-GHZ state analysis.}
\end{center}
\end{figure}
Before we start this protocol, we first transform the states $|\Phi^{\pm}_{1}\rangle_{N,m}$, $|\Phi^{\pm}_{2}\rangle_{N,m}$, $\cdots$, $|\Phi^{\pm}_{2^{N-1}}\rangle_{N,m}$ to $|\Phi^{\pm}_{1}\rangle_{N,2}$, $|\Phi^{\pm}_{2}\rangle_{N,2}$,  $\cdots$, $|\Phi^{\pm}_{2^{N-1}}\rangle_{N,2}$, respectively. Such transformation can be completed by performing $N-2$ Hadamard operations on each qubit and measuring these $N-2$ qubits in $\{|0\rangle, |1\rangle\}$ basis. From the measurement results, if the number of  $|1\rangle$ is even, the states $|\Phi^{\pm}_{1}\rangle_{N,m}$, $|\Phi^{\pm}_{2}\rangle_{N,m}$, $\cdots$, $|\Phi^{\pm}_{2^{N-1}}\rangle_{N,m}$ have fully transformed to $|\Phi^{\pm}_{1}\rangle_{N,2}$, $|\Phi^{\pm}_{2}\rangle_{N,2}$,  $\cdots$, $|\Phi^{\pm}_{2^{N-1}}\rangle_{N,2}$, respectively. Otherwise, if the number of  $|1\rangle$ is odd,
the states $|\Phi^{\pm}_{1}\rangle_{N,m}$, $|\Phi^{\pm}_{2}\rangle_{N,m}$, $\cdots$, $|\Phi^{\pm}_{2^{N-1}}\rangle_{N,m}$ have fully transformed to $|\Phi^{\mp}_{1}\rangle_{N,2}$, $|\Phi^{\mp}_{2}\rangle_{N,2}$,  $\cdots$, $|\Phi^{\mp}_{2^{N-1}}\rangle_{N,2}$, respectively. In this way, we should perform a phase-flip operation to transform the states $|\Phi^{\mp}_{1}\rangle_{N,2}$, $|\Phi^{\mp}_{2}\rangle_{N,2}$,  $\cdots$, $|\Phi^{\mp}_{2^{N-1}}\rangle_{N,2}$ to   $|\Phi^{\pm}_{1}\rangle_{N,2}$, $|\Phi^{\pm}_{2}\rangle_{N,2}$,  $\cdots$, $|\Phi^{\pm}_{2^{N-1}}\rangle_{N,2}$, respectively.
In Fig. 2, we denote each logic qubit $|\phi^{\pm}\rangle$ as $A$, $B$, $\cdots$, etc.  We first perform the Hadamard operation on each physical qubit, and make the states in Eq. (\ref{multi2}) becomes
\begin{eqnarray}
&&|\Phi^{\pm}_{1}\rangle_{N,2}=\frac{1}{\sqrt{2}}(|\phi^{+}\rangle^{\otimes N}\pm|\psi^{+}\rangle^{\otimes N}),\nonumber\\
&&|\Phi^{\pm}_{2}\rangle_{N,2}=\frac{1}{\sqrt{2}}(|\psi^{+}\rangle|\phi^{+}\rangle^{\otimes N-1}\pm|\phi^{+}\rangle|\psi^{+}\rangle^{\otimes N-1}),\nonumber\\
&&\cdots\nonumber\\
&&|\Phi^{\pm}_{2^{N-1}}\rangle_{N,2}=\frac{1}{\sqrt{2}}(|\phi^{+}\rangle^{\otimes N-1}|\psi^{+}\rangle\pm|\psi^{+}\rangle^{\otimes N-1}|\phi^{+}\rangle.\nonumber\\\label{multi3}
\end{eqnarray}
After passing through the CNOT gates, the states in Eq. (\ref{multi3}) can be written as
\begin{eqnarray}
|\Phi^{\pm}_{1}\rangle_{N,2}&\rightarrow&\frac{1}{\sqrt{2}}(|+\rangle^{\otimes N}|0\rangle^{\otimes N}\pm|+\rangle^{\otimes N}|1\rangle^{\otimes N}),\nonumber\\
|\Phi^{\pm}_{2}\rangle_{N,2}&\rightarrow&\frac{1}{\sqrt{2}}(|+\rangle^{\otimes N}|1\rangle|0\rangle^{\otimes N-1}\nonumber\\
&\pm&|+\rangle^{\otimes N}|0\rangle|1\rangle^{\otimes N-1})\nonumber\\
&\cdots&\nonumber\\
|\Phi^{\pm}_{2^{N-1}}\rangle_{N,2}&\rightarrow&\frac{1}{\sqrt{2}}(|+\rangle^{\otimes N}|0\rangle^{\otimes N-1}|1\rangle\nonumber\\
&\pm&|+\rangle^{\otimes N}|1\rangle^{\otimes N-1}|0\rangle).\label{multi4}
\end{eqnarray}
From Eq. (\ref{multi4}), all the first physical qubits a$_{1}$, b$_{1}$ $\cdots$, n$_{1}$ in each logic qubits disentangle with the second one. By removing
these qubits, the states in Eq. (\ref{multi4}) can be written as the standard $N$-particle GHZ state of the form
\begin{eqnarray}
&&|\Phi^{\pm}_{1}\rangle_{N,2}\rightarrow|\Phi^{\pm}_{1}\rangle_{N}=\frac{1}{\sqrt{2}}(|0\rangle^{\otimes N}\pm|1\rangle^{\otimes N}),\nonumber\\
&&|\Phi^{\pm}_{2}\rangle_{N,2}\rightarrow|\Phi^{\pm}_{2}\rangle_{N}=\frac{1}{\sqrt{2}}(|1\rangle|0\rangle^{\otimes N-1}\pm|0\rangle|1\rangle^{\otimes N-1}),\nonumber\\
&&\cdots\nonumber\\
&&|\Phi^{\pm}_{2^{N-1}}\rangle_{N,2}\rightarrow|\Phi^{\pm}_{2^{N}}\rangle_{N}=\frac{1}{\sqrt{2}}(|0\rangle^{\otimes N-1}|1\rangle\pm|1\rangle^{\otimes N-1}|0\rangle).\nonumber\\\label{multi5}
\end{eqnarray}
The $N$ particles are denoted as a$_{2}$, b$_{2}$, $\cdots$, n$_{2}$, as shown in Fig. 2. Therefore, the discrimination of the  C-GHZ state equals to the discrimination of the standard $N$-particle GHZ state. The next step can be described as follows.
We first perform the CNOT gate between the neighboring qubits (n-1)$_{2}$ and n$_{2}$. In Fig. 2, we denote $k\equiv n-1$. The qubit k$_{2}$ is the source qubit and the qubit n$_{2}$ is the target qubit. After performing the CNOT operation, we let the qubit (n-2)$_{2}$ be the source qubit and the qubit k$_{2}$ ((n-1)$_{2}$) be the target qubit and perform the CNOT operation again. In each round, we let the l$_{2}$ (l$_{2}$=1,2,$\cdots$, N-1) qubit  be the source qubit and the (l+1)$_{2}$ qubit be the target qubit and perform the CNOT operation. After performing  $N-1$ CNOT operations, the states in Eq. (\ref{multi5}) become
\begin{eqnarray}
|\Phi^{\pm}_{1}\rangle_{N}&\rightarrow&\frac{1}{\sqrt{2}}|\pm\rangle|0\rangle^{\otimes N-1},\nonumber\\
|\Phi^{\pm}_{2}\rangle_{N}&\rightarrow&\frac{1}{\sqrt{2}}|\pm\rangle|1\rangle|0\rangle^{\otimes N-2},\nonumber\\
&&\cdots\nonumber\\
|\Phi^{\pm}_{2^{N}}\rangle_{N}&\rightarrow&\frac{1}{\sqrt{2}}|\pm\rangle|0\rangle^{\otimes N-2}|1\rangle.
\end{eqnarray}
Finally, after performing the Hadmard operation on the first qubit a$_{2}$ to transform $|+\rangle$ to $|0\rangle$ and  $|-\rangle$ to $|1\rangle$, all the
states in Eq. (\ref{multi1}) can be distinguished deterministically by measuring all the qubits in the basis $\{|0\rangle,|1\rangle\}$.
For example, if the measurement results are $|0\rangle|0\rangle\cdots|0\rangle$, the original state is $|\Phi^{+}_{1}\rangle_{N,m}$.
If the  measurement results are $|1\rangle|0\rangle\cdots|0\rangle$, the original state is $|\Phi^{-}_{1}\rangle_{N,m}$. In this way, we can distinguish
arbitrary C-GHZ state.

\section{Discussion}
So far, we have fully explained the LBSA and the  C-GHZ state analysis. It is interesting to discuss some applications of such state analysis. Quantum teleportation \cite{teleportation} and entanglement swapping \cite{repeater} are two unique techniques in quantum communication. The former can transmit an unknown state of the information encoded in a particle to a remote point without distributing the particle itself. The later can be used to extend the distance of quantum communication. We will show that the unknown logic qubit can also be teleportated and entanglement swapping based on the logic entanglement can also be performed in principle.

\subsection{Logic qubit teleportation}
Suppose that  an arbitrary logic qubit in Alice's laboratory can be defined as
\begin{eqnarray}
|\varphi\rangle_{A}=\alpha|GHZ^{+}_{m}\rangle_{A}+\beta|GHZ^{-}_{m}\rangle_{A},
\end{eqnarray}
with $|\alpha|^{2}+|\beta|^{2}=1$. Alice and Bob share the logic qubits entanglement in the channel BC of the form
\begin{eqnarray}
|\Psi\rangle_{BC}&=&\frac{1}{\sqrt{2}}(|GHZ^{+}_{m}\rangle_{B}|GHZ^{+}_{m}\rangle_{C}\nonumber\\
&+&|GHZ^{-}_{m}\rangle_{B}|GHZ^{-}_{m}\rangle_{C}).
\end{eqnarray}
Alice performs the LBSA on her two logic qubits A and B. The whole state can be described as
\begin{eqnarray}
&&|\varphi\rangle_{A}\otimes|\Psi\rangle_{BC}=(\alpha|GHZ^{+}_{m}\rangle_{A}+\beta|GHZ^{-}_{m}\rangle_{A})\nonumber\\
&&\otimes[\frac{1}{\sqrt{2}}(|GHZ^{+}_{m}\rangle_{B}|GHZ^{+}_{m}\rangle_{C}
+|GHZ^{-}_{m}\rangle_{B}|GHZ^{-}_{m}\rangle_{C})]\nonumber\\
&=&\frac{1}{2}[\frac{1}{\sqrt{2}}(|GHZ^{+}_{m}\rangle_{A}|GHZ^{+}_{m}\rangle_{B}
+|GHZ^{-}_{m}\rangle_{A}|GHZ^{-}_{m}\rangle_{B})]\nonumber\\
&\otimes&(\alpha|GHZ^{+}_{m}\rangle_{C}+\beta|GHZ^{-}_{m}\rangle_{C})\nonumber\\
&+&\frac{1}{2}[\frac{1}{\sqrt{2}}(|GHZ^{+}_{m}\rangle_{A}|GHZ^{+}_{m}\rangle_{B}
-|GHZ^{-}_{m}\rangle_{A}|GHZ^{-}_{m}\rangle_{B})]\nonumber\\
&\otimes&(\alpha|GHZ^{+}_{m}\rangle_{C}-\beta|GHZ^{-}_{m}\rangle_{C})\nonumber\\
&+&\frac{1}{2}[\frac{1}{\sqrt{2}}(|GHZ^{+}_{m}\rangle_{A}|GHZ^{-}_{m}\rangle_{B}
+|GHZ^{-}_{m}\rangle_{A}|GHZ^{+}_{m}\rangle_{B})]\nonumber\\
&\otimes&(\alpha|GHZ^{-}_{m}\rangle_{C}+\beta|GHZ^{+}_{m}\rangle_{C})\nonumber\\
&+&\frac{1}{2}[\frac{1}{\sqrt{2}}(|GHZ^{+}_{m}\rangle_{A}|GHZ^{-}_{m}\rangle_{B}
-|GHZ^{-}_{m}\rangle_{A}|GHZ^{+}_{m}\rangle_{B})]\nonumber\\
&\otimes&(\alpha|GHZ^{-}_{m}\rangle_{C}-\beta|GHZ^{+}_{m}\rangle_{C}).\label{teleportation}
\end{eqnarray}
Obviously, from Eq. (\ref{teleportation}), according to the LBSA, Alice can teleportate the arbitrary
logic qubit $|\varphi\rangle_{A}$ to Bob.

\subsection{Logic qubit entanglement swapping}
Let pairs AB and CD be in the following logic entangled
states
\begin{eqnarray}
|\Psi\rangle_{AB}&=&\frac{1}{\sqrt{2}}(|GHZ^{+}_{m}\rangle_{A}|GHZ^{+}_{m}\rangle_{B}\nonumber\\
&+&|GHZ^{-}_{m}\rangle_{A}|GHZ^{-}_{m}\rangle_{B}),
\end{eqnarray}
and
\begin{eqnarray}
|\Psi\rangle_{CD}&=&\frac{1}{\sqrt{2}}(|GHZ^{+}_{m}\rangle_{C}|GHZ^{+}_{m}\rangle_{D}\nonumber\\
&+&|GHZ^{-}_{m}\rangle_{C}|GHZ^{-}_{m}\rangle_{D}).
\end{eqnarray}
If we perform a logic Bell-state measurement between the logic qubits B and C, the whole system can be described as
\begin{eqnarray}
&&|\Psi\rangle_{AB}\otimes|\Psi\rangle_{CD}=[\frac{1}{\sqrt{2}}(|GHZ^{+}_{m}\rangle_{A}|GHZ^{+}_{m}\rangle_{B}\nonumber\\
&+&|GHZ^{-}_{m}\rangle_{A}|GHZ^{-}_{m}\rangle_{B})]
\otimes[\frac{1}{\sqrt{2}}(|GHZ^{+}_{m}\rangle_{C}|GHZ^{+}_{m}\rangle_{D}\nonumber\\
&+&|GHZ^{-}_{m}\rangle_{C}|GHZ^{-}_{m}\rangle_{D})]\nonumber\\
&&=\frac{1}{2}[\frac{1}{\sqrt{2}}(|GHZ^{+}_{m}\rangle_{A}|GHZ^{+}_{m}\rangle_{D}\nonumber\\
&+&|GHZ^{-}_{m}\rangle_{A}|GHZ^{-}_{m}\rangle_{D})]
\otimes[\frac{1}{\sqrt{2}}(|GHZ^{+}_{m}\rangle_{B}|GHZ^{+}_{m}\rangle_{C}\nonumber\\
&+&|GHZ^{-}_{m}\rangle_{B}|GHZ^{-}_{m}\rangle_{C})]\nonumber\\
&+&\frac{1}{2}[\frac{1}{\sqrt{2}}(|GHZ^{+}_{m}\rangle_{A}|GHZ^{+}_{m}\rangle_{D}\nonumber\\
&-&|GHZ^{-}_{m}\rangle_{A}|GHZ^{-}_{m}\rangle_{D})]
\otimes[\frac{1}{\sqrt{2}}(|GHZ^{+}_{m}\rangle_{B}|GHZ^{+}_{m}\rangle_{C}\nonumber\\
&-&|GHZ^{-}_{m}\rangle_{B}|GHZ^{-}_{m}\rangle_{C})]\nonumber\\
&+&\frac{1}{2}[\frac{1}{\sqrt{2}}(|GHZ^{+}_{m}\rangle_{A}|GHZ^{-}_{m}\rangle_{D}\nonumber\\
&+&|GHZ^{-}_{m}\rangle_{A}|GHZ^{+}_{m}\rangle_{D})]
\otimes[\frac{1}{\sqrt{2}}(|GHZ^{+}_{m}\rangle_{B}|GHZ^{-}_{m}\rangle_{C}\nonumber\\
&+&|GHZ^{-}_{m}\rangle_{B}|GHZ^{+}_{m}\rangle_{C})]\nonumber\\
&+&\frac{1}{2}[\frac{1}{\sqrt{2}}(|GHZ^{+}_{m}\rangle_{A}|GHZ^{-}_{m}\rangle_{D}\nonumber\\
&-&|GHZ^{-}_{m}\rangle_{A}|GHZ^{+}_{m}\rangle_{D})]
\otimes[\frac{1}{\sqrt{2}}(|GHZ^{+}_{m}\rangle_{B}|GHZ^{-}_{m}\rangle_{C}\nonumber\\
&-&|GHZ^{-}_{m}\rangle_{B}|GHZ^{+}_{m}\rangle_{C})].\label{swapping}
\end{eqnarray}
From Eq. (\ref{swapping}), if the LBSA between B and C is projected to $\frac{1}{\sqrt{2}}(|GHZ^{+}_{m}\rangle_{B}|GHZ^{+}_{m}\rangle_{C}
+|GHZ^{-}_{m}\rangle_{B}|GHZ^{-}_{m}\rangle_{C})$, the logic qubits A and D will collapse to the same state with
$\frac{1}{\sqrt{2}}(|GHZ^{+}_{m}\rangle_{A}|GHZ^{+}_{m}\rangle_{D}+|GHZ^{-}_{m}\rangle_{A}|GHZ^{-}_{m}\rangle_{D})$, with the probability of $\frac{1}{4}$. On the other hand, if the
measurement results of B and C are the other states, we can also obtain the  corresponding entangled states, which can be transformed to
$|\Psi\rangle_{AD}$ deterministically.

\section{Conclusion}
In conclusion, we have presented a protocol for LBSA. We exploit the CNOT gate, Hadamard gate and single qubit measurement
to complete the task. We also showed that arbitrary C-GHZ state can also be well distinguished.
It is shown that the number of physical qubit in each logic qubit does not affect the analysis of C-GHZ state. Therefore, the analysis of
C-GHZ state with arbitrary $N$ and $m$ equals to the analysis of the C-GHZ state with $m=2$. Both the LBSA and C-GHZ state can be simplified to the
standard Bell-state analysis and GHZ state analysis, respectively, which can be well distinguished with the help of CNOT gate.
As the C-GHZ state is more robust than the normal GHZ states in a noisy environment \cite{cghz,pan}, our LBSA shows that it is possible to perform the
long-distance quantum communication based on the logic qubits rather than the physical qubits directly.

\section*{ACKNOWLEDGEMENTS}
This work is supported by the National Natural Science Foundation of
China under Grant No. 11104159 and 11347110,  and a Project Funded by the Priority
Academic Program Development of Jiangsu Higher Education
Institutions.

\end{document}